\newcommand{\sign}{{\rm sgn}\,}

\documentclass[%
 reprint,
superscriptaddress,
 amsmath,amssymb,
 aps,
]{revtex4-2}

\def\XXint#1#2#3{{\setbox0=\hbox{$#1{#2#3}{\int}$}
     \vcenter{\hbox{$#2#3$}}\kern-.5\wd0}}

\usepackage{amsthm} 
\usepackage{tikz}
\usepackage{pgfplots}
\usepackage{color}
\usepackage{comment}
\usepackage{soul}
\newtheorem{property}{Property}
\newtheorem{condition}{Condition}
\def\doi{http://dx.doi.org/}
\usepackage{braket}
\definecolor{OliveGreen}{RGB}{85,107,47}
\definecolor{NavyBlue}{RGB}{0,0,128}
\usepackage[colorlinks,bookmarks=false,citecolor=NavyBlue,linkcolor=OliveGreen,urlcolor=blue]{hyperref}
\usepackage{feynmf,tikz, relsize}
\newcommand{\bw}{\begin{widetext}}
\newcommand{\ew}{\end{widetext}}
\newcommand{\be}{\begin{equation}}
\newcommand{\ee}{\end{equation}}
\newcommand{\bea}{\begin{eqnarray}}
\newcommand{\eea}{\end{eqnarray}}
\def\nn{\nonumber\\}
\def\fr#1{(\ref{#1})}

\usepackage{graphicx}
\usepackage{dcolumn}
\usepackage{bm}
\pgfplotsset{compat=1.17}

\usepackage{xr} 
\begin{document}

\title{BBGKY Hierarchy and Generalised Hydrodynamics}

\author{Bruno Bertini}
\affiliation{School of Physics and Astronomy, University of Nottingham, Nottingham, NG7 2RD, UK}

\author{Fabian H.L. Essler}
\affiliation{Rudolf Peierls Centre for Theoretical Physics, Clarendon Laboratory, Oxford OX1 3PU, UK}

\author{Etienne Granet}
\affiliation{Kadanoff Center for Theoretical Physics, University of Chicago, 5640 South Ellis Ave., Chicago, IL 60637, USA}

\date{\today}

\begin{abstract}
We consider fermions defined on a continuous one-dimensional interval and subject to weak repulsive two-body interactions. We show that it is possible to perturbatively construct an extensive number of mutually compatible conserved charges for any interaction potential. However, the contributions to the densities of these charges at second order and higher are generally non-local and become spatially localized only if the potential fulfils certain compatibility conditions. We prove that the only solutions to the first of these conditions are the Cheon-Shigehara potential (fermionic dual to the Lieb-Liniger model) and the Calogero-Sutherland potentials. We use our construction to show how Generalized Hydrodynamics (GHD) emerges from the Bogoliubov--Born--Green--Kirkwood--Yvon hierarchy, and argue that GHD in the weak interaction regime is robust under non-integrable perturbations.
\end{abstract}

\maketitle

Finding an efficient description for the non-equilibrium dynamics of quantum many-particle systems has been a key problem in theoretical physics since the birth of quantum mechanics~\cite{vonneumann2010proof}. During the last two decades there has been an upsurge in interest as a result of significant experimental advances ~\cite{bloch2008many} and potential technological applications~\cite{arute2019quantum}. In spite of remarkable progress in our understanding of out-of-equilibrium quantum matter~\cite{polkovnikov2011colloquium, rigol2008thermalization, gogolin2016equilibration, calabrese2016introduction, serbyn2021quantum, bertini2021finite, bastianello2021introduction}, however, an efficient and accurate description of its dynamics remains out of reach. 

In essence, what makes this problem so hard is the lack of general methods to tackle it. Exact methods are restricted to a small number of fine-tuned many-body systems 
~\cite{bertini2019entanglement, piroli2020exact, klobas2021exact, klobas2021exact2, granet2022out} and there is currently no controlled approximation scheme applicable to generic interacting systems.
Numerical methods are typically limited by the accessible system sizes~\cite{rigol2008thermalization, carleo2012localization} or by the class of initial states that can be accommodated~\cite{mallayya2018quantum}. Even in one dimension, where techniques based on matrix product states~\cite{daley2004time, white2004real, schollwoeck2011the} give access to large systems, their applicability is limited to short times by the rapid growth of quantum entanglement. Furthermore, continuum models --- which describe many relevant experiments --- provide additional obstacles to numerical approaches.

For weakly interacting systems the situation is substantially simpler because, at least in principle, a general description of the dynamics can be attained by using the celebrated Bogoliubov--Born--Green--Kirkwood--Yvon (BBGKY) hierarchy~\cite{bogoliubov1970lectures,bonitz2015quantum,huang2008statistical}, which encodes the Heisenberg equations of motion for the reduced density matrices. To illustrate it let us consider for simplicity and concreteness a system of spinless fermions: in this case the reduced density matrices are written as 
\be
\rho_n({\bf x},{\bf y};t)={\rm Tr}\left[\rho(t)\psi^\dagger_{x_1}\dots \psi^\dagger_{x_n}\psi^{\phantom{\dagger}}_{y_n}\dots \psi^{\phantom{\dagger}}_{y_1}\right],
\ee
where $\psi^\dagger_x$ and $\psi_x$ are fermionic creation and annihilation operators and $\rho(t)$ denotes the
density matrix specifying the state of the system at time $t$. Assuming that the fermions interact via a two-body potential $V(x-y)$ the BBGKY hierarchy takes the form
\begin{align}
&i\partial_t \rho_n({\bf x},{\bf y};t)-(H^{(n)}_{\bf x}-H^{(n)}_{\bf y})\rho_n({\bf x},{\bf y};t)
=\label{BBGKY}\\
&\sum_{j=1}^n\int dw [V(x_j-w)-V(y_j-w)]\rho_{n+1}({\bf x},w,{\bf y},w;t),\notag
\end{align}
where $H^{(n)}_{\bf x}$ is the first-quantized $n$-particle Hamiltonian in the position representation. In principle the system (\ref{BBGKY}) gives a complete account of the non-equilibrium dynamics of many-particle quantum systems. In practice it is necessary to truncate it, which can be justified e.g. for weak interactions. Retaining only two-particle cumulants gives rise \cite{erdos2004quantum} to the ubiquitous quantum Boltzmann equation (QBE) for the Wigner function
\be
f_t(x,p)=\int dz e^{-ipz}\ {\rm Tr}\big[\rho(t)\psi^\dagger_{x+\frac{z}{2}}\psi^{\phantom{\dagger}}_{x-\frac{z}{2}}\big].\label{wigner}
\ee
The QBE holds in the regime of weak spatial variations and large times, i.e.\ the Euler scaling limit \cite{erdos2004quantum}, and under the assumption of local relaxation can be further reduced to a set of hydrodynamic equations, which are obtained from the continuity equations of particle number (or mass), energy, and momentum~\cite{huang2008statistical,bonitz2015quantum}.

The situation is very different, and significantly richer, in quantum integrable systems~\cite{korepin1993quantum, sutherland2004beautiful}, which are characterized by having an extensive number of (mutually compatible) conservation laws with densities that are sufficiently local in space. These conservation laws give rise to the existence of stable quasi-particles over macro states at finite energy densities~\footnote{In the following we restrict our discussion to the simplest case where there is only a single species of quasi-particles.}. Integrable models do not thermalize but relax to a much wider class of equilibrium states known as Generalized Gibbs Ensembles~\cite{essler2016quench, vidmar2016generalized}, which can be fully characterized in terms of their respective quasi-particle density in momentum space $\rho(k)$. In these systems the dynamics of local observables close to local equilibrium is described by GHD~\cite{bertini2016transport, castro2016emergent}, which can be expressed as an evolution equation for a space and time dependent density $\rho_{x,t}(k)$ of the stable quasi-particles
\begin{equation}\label{ghd}
    \partial_t \rho_{x,t}(k)+\partial_x(v_{x,t}(k)\rho_{x,t}(k))=0\ .
\end{equation}
Here $v_{x,t}(k)$ is a (known) quasi-particle velocity that depends on $\rho_{x,t}(\cdot)$. GHD is obtained from the system of continuity equations for the extensive number of conservation laws, postulating local relaxation to an equilibrium state with quasi-particle density $\rho_{x,t}(k)$, and then inferring its evolution equation (\ref{ghd}). The GHD equation (\ref{ghd}) can be viewed as a kinetic theory governed by a \emph{dissipationless} Boltzmann Equation~\cite{bertini2016transport} where the velocity is a non-linear functional of the density itself~\cite{percus1969exact, boldrighini1983one}. Hence it suggests the existence of an operator $n(x,k)$ expressed in terms of fermions $\psi_x$ localized near $x$, whose expectation value is $\rho_{x,t}(k)$, and for which the BBGKY hierarchy would re-organize into a dissipationless QBE. Such an operator, however, is only known in the non-interacting case, where it is given by the Wigner operator (whose expectation value is (\ref{wigner})).

In this letter we show that under certain conditions the BBGKY hierarchy can be reformulated as GHD equations. We establish that, surprisingly, one can construct conserved charges for \emph{arbitrary local} but weak interaction potentials by appropriately dressing the non-interacting modes. Crucially, however, these conserved charges are generally non-local and acquire good locality properties only for integrable models. This means that even though it is possible to explicitly construct operators fulfilling a dissipationless QBE for generic interacting models, these operators are non-local and the equation cannot be used to infer the dynamics of physical observables. In contrast, in the integrable case the evolution equation reduces to the GHD equation for expectation values of local mode occupation numbers. 
As discussed in the following, this also gives a simple operational criterion to assess the integrability of a given interaction potential.

In the remainder of this letter we consider a system of interacting fermions on a ring of size $L$ with Hamiltonian
$H=H_0+H_1$
\be
H_0=\sum_{p} p^2\psi_p^\dag \psi_{p}\ ,\
H_1=\frac{1}{L}\sum_{{\boldsymbol p}} V({\boldsymbol p}) \psi_{p_1}^\dag \psi_{p_2}^\dag\psi^{\phantom{\dag}}_{p_3}\psi^{\phantom{\dag}}_{p_4},
\label{eq:FermionicH}
\ee
where all summations are on ``free momenta" $p={2\pi n}/{L}$ with $n\in\mathbb{Z}$ and where $\psi_p^\dagger$ and $\psi_p^{\phantom{\dagger}}$ are fermionic creation and annihilation operators with canonical anti commutation relations ${\{\psi_{p_1}^\dag, \psi^\dag_{p_2}\}=0=\{\psi_{p_1},\psi_{p_2}\}}$ and $\{\psi_{p_1}^\dag, \psi_{p_2}\}=\delta_{p_1,p_2}$. Here $V({\bf p})$ is a two-particle interaction potential, which by virtue of the anti-commutation relations can be cast in the form
\be
V(\boldsymbol p) =  \frac{1}{4}\delta_{p_1+p_2,p_3+p_4} \mathcal A_{p_1 p_2}\mathcal A_{p_3 p_4}[V(p_1-p_3)],
\label{eq:VW}
\ee
where $\mathcal A_{k_1 \cdots k_n}$ is an operator that acts by anti-symmetrising in $\{k_1, \ldots, k_n\}$. Without loss of generality we take $V(k)$ to be even in $k$ and $V(0)=0$~\footnote{$V(k)$ is always even for identical particles. However the assumption $V(0)=0$ holds only for fermions. A non-zero $V(0)$ corresponds to a $\delta$ interaction in real space, which is known to have no effects on fermions.}.

In particular, the choice 
\be
V(k)=E_a(k):= -\frac{\beta}{a^2}\int {\rm d}x\,\,  \frac{\sigma''(x/a) (e^{i k x}-1)}{x+\beta\sigma(x/a)}  \,,
\label{eq:Etiennespotential}
\ee
where $\sigma(x)$ is a smooth odd function with ${\sigma'(x)\geq 0}$, ${\sigma'(0)> 0}$, ${\lim_{x\to\infty}\sigma(x)=1}$, ${\lim_{x\to\infty} x^2 \sigma''(x)=0}$, produces the second-quantised (and regularised) Cheon Shigehara model of Ref.~\cite{granet2021realizing,granet2022regularisation}, i.e., an integrable fermionic dual of the Lieb-Liniger model \cite{lieb1963exact}. We stress that the regularisation~\eqref{eq:Etiennespotential} allows for a perturbative expansion at second order without renormalisation.

In order to obtain GHD we follow the logic of \cite{bertini2016transport,castro2016emergent}, but aim to explicitly construct the conserved charges from the BBGKY hierarchy. These are derived by assuming that the potential $V(k)$ can be treated perturbatively. 
We note that for the potential (\ref{eq:Etiennespotential}) perturbation theory is believed to have a finite radius of convergence, \emph{cf.} \cite{lang2019conjectures,granet2020a,granet2021systematic}.

Our first step is to find all the conserved charges of \eqref{eq:FermionicH}. We begin by considering an operator $Q$ and assuming that it has a regular perturbative expansion 
\be
Q= \sum_{m=0}^\infty Q_{m},
\ee
where $Q_{m}$ is of order $m$ in $V(k)$. $Q$ is conserved if the operators $Q_{m}$ fulfil  
\be
[H_0, Q_{m}]  = - [H_1, Q_{m-1}],
\label{eq:chargePT}
\ee
where we set $Q_{-1}=0$. In addition, we restrict our discussion to cases where the particle number operator $N=\sum_p \psi_p^\dag\psi_p$ is part of the tower of conserved charges. This implies that all $Q_n$ are expressed as sums of monomials involving equal numbers of $\psi_p^\dag$ and $\psi_{q}$. It is easy to see that the zeroth order of \eqref{eq:chargePT} is solved by the following number-conserving charges 
\be
Q_{f;0} =  \sum_{p} f(p) \psi_p^\dag \psi_{p},
\label{eq:zerothQ}
\ee
where $f(p)$ is any function, which we take to be smooth. We note that charges corresponding to different $f$s commute, but \eqref{eq:zerothQ} is clearly not the most general choice for a conserved charge of $H_0$. Any energy-diagonal operator (i.e.\ any operator that is diagonal in an eigenbasis of $H_0$) could be taken as a conserved charge of $H_0$, with any $V$-dependent prefactor appearing at higher orders. The densities of such more complicated energy-diagonal operators, however, are generically non-local in real space. We therefore restrict our search to zeroth-order charges of the form~\eqref{eq:zerothQ} and leave out energy-diagonal contributions at higher orders. 
We expect this ``minimal ansatz" to be sufficient for the perturbative construction of complete sets of conserved charges in all integrable models of the form \eqref{eq:FermionicH} featuring a single species of quasi-particles, i.e. cases in which
the stable quasi-particles can be thought of as being ``adiabatically connected" to the free fermions $\psi_p^\dagger\psi_p^{\phantom{\dagger}}$. 

Starting from \eqref{eq:zerothQ} we can directly use \eqref{eq:chargePT} and our minimal ansatz to recursively generate the higher orders of $Q_{f}$. In particular, for $m=1$ the equation can be solved for any potential giving~\footnote{See the Supplemental Material for: (i) an explicit derivation of \eqref{eq:Qf1st} and \eqref{eq:Qf2nd}; (ii) a proof of Property~\ref{p:prop1}; (iii) a proof that the charges are in involution; (iv) an explicit expression of the current $\hat j_{f,x}(t)$ at first order; (v) the derivation of \eqref{eq:pertcharge} and \eqref{eq:pertcurr} from Thermodynamic Bethe ansatz in the cases of Lieb Liniger and Calogero Sutherland.}
\be
Q_{f;1}  = \frac{1}{L}\!\sum_{{\boldsymbol k}} g_{f;1}^{(4)}({\boldsymbol k}) \psi_{k_1}^\dag \psi^{\dag}_{k_2}\psi_{k_3}^{\phantom{\dag}} \psi^{\phantom{\dag}}_{k_1+k_2-k_3}, 
\label{eq:Qf1st}
\ee 
where the function $g_{f;1}^{(4)}({\boldsymbol k})$ is non-singular for any choice of $V(k)$, and has the same regularity as $V(k)$: its explicit expression is given in \eqref{eq:g1}. 

In contrast, for $m=2$ Eq.~\eqref{eq:chargePT} does not always admit a solution in the framework of our minimal ansatz, i.e.
starting from the zeroth order \eqref{eq:zerothQ} and omitting energy-diagonal contributions at higher orders. This is because $[H_1, Q_{f;1}]$ generically contains an energy-diagonal component and therefore cannot be expressed in the form $[H_0, Q_{f;2}]$. Indeed, the energy-diagonal part of the latter commutator is always zero. In particular, defining $\mathcal{S}_6:=\mathcal M_{6} \setminus \mathcal N_{6}$ with    
\begin{align}
\mathcal M_{6} &:=\! 
 \Bigl\{{k}_j\!:\,\, \sum_{j=1}^3  k_j^2=\sum_{j=4}^6  k_j^2,\,\,\, \sum_{j=1}^3 k_j\!=\!\sum_{j=4}^6 k_j\Bigr\}\ ,\label{eq:M6}\\
\mathcal N_{6} &:=\! 
 \Bigl\{{k}_j\!:\,\, \{k_1,k_2,k_3\}=\{k_4,k_5,k_6\}
\Bigr\},
\end{align}
we find the following solvability condition
\begin{condition}
\label{eq:condition1}
For $Q_{f;1}$ given in Eq.~\eqref{eq:Qf1st}, Eq.~\eqref{eq:chargePT} admits solution for $m=2$ and all $f$ only if 
\be
\mathcal A_{k_1k_2k_3}\mathcal A_{k_4k_5k_6}\!\!\left[\frac{V(k_4-k_1)V(k_3-k_5)}{(k_{1}-k_{4})(k_2-k_4)}\right] = 0,
\label{eq:keyproperty}
\ee
for every  $\boldsymbol k\in \mathcal{S}_6$.  
\end{condition}
\noindent This condition can be shown to be equivalent to the request that the connected 2nd order contribution to the three-particle S-matrix vanishes for different sets of incoming and outgoing momenta~\cite{bertini2022wip}.

If Condition~\ref{eq:condition1} is fulfilled the second order charge can be expressed as~\cite{Note2} 
\begin{align}
Q_{f;2} &= \frac{1}{L}\! \sum_{{\boldsymbol k}} g_{f;2}^{(4)}({\boldsymbol k}) \psi_{k_1}^\dag \psi^{\dag}_{k_2}\psi_{k_3}^{\phantom{\dag}} \psi^{\phantom{\dag}}_{k_1+k_2-k_3} \label{eq:Qf2nd}\\
&+\!\!\frac{1}{L^2}\! \sum_{{\boldsymbol k}} g_{f;2}^{(6)}({\boldsymbol k}) \psi^{\dag}_{k_1}\psi^{\dag}_{k_2} \psi^{\dag}_{k_3} \psi^{\phantom{\dag}}_{k_4} \psi^{\phantom{\dag}}_{k_5} \psi^{\phantom{\dag}}_{k_1+k_2+k_3- k_4-k_5}\!,\notag
\end{align}
where $\{g_{f;2}^{(n)}({\boldsymbol k})\}_{n=4,6}$ are regular and their explicit expressions are given in Eqs.~\eqref{eq:g24} and \eqref{eq:g26}.

Proceeding in this way we obtain a set of charges
\be
\{Q_{f} = Q_{f;0}+Q_{f;1}+Q_{f;2}, \quad \text{any smooth}\quad f(k)\}\ ,
\label{eq:charges}
\ee
conserved up to $O(V^3)$. Our construction can be readily extended to higher orders. 
Crucially the charges~\eqref{eq:charges} are mutually compatible (it is shown in \cite{Note2} that they commute up to $O(V^3)$) and for smooth $f(k)$ and $V(k)$ they are quasi-local, i.e.\ their density 
is exponentially localised. The latter property can be shown by expressing $Q_{f;m}$ in terms of real space fermions $\psi(x)= \sum_p \psi_p e^{i x p}/{\sqrt L}$. For instance, considering \eqref{eq:Qf1st} we have  
\be
\!\!\! Q_{f;1} \!\! = \!\!\int \!\!{\rm d}  \boldsymbol x\, C_{f;1}(x_1-x_4,\ldots,x_3-x_4)\, \psi^\dag_{x_1} \!\!\ldots \psi_{x_4},
\label{eq:Qf1stpos}
 \ee
 where 
 \be
 C_{f;1}(y_1,y_2,y_3)=\frac{1}{L^3} \sum_{{\boldsymbol k}}  g_{f;1}^{(4)}({\boldsymbol k}) e^{i (k_1 y_1+ k_2 y_2 - k_3 y_3)}.
 \ee 
In the thermodynamic limit $C_{f;1}(\bf x)$ becomes the Fourier transform of $g_{f;1}^{(4)}({\boldsymbol k})$, which is smooth for $f(k)$ and $V(k)$ smooth. Therefore, quasi-locality is guaranteed by standard Fourier analysis~\cite{strichartz2003guide}. If $V(k)$ is regular but not smooth, the densities still decay in real space, but generically as power-laws.

If Condition~\ref{eq:condition1} is not fulfilled one can construct charges at second order by either: (i) adding appropriate energy-diagonal contributions to $Q_{f,0}$ (or $Q_{f;1}$) to subtract the energy diagonal part of $[H_1, Q_{f,1}]$~\footnote{This calculation will be reported elsewhere \cite{bertini2022wip}.}; (ii) appropriately deforming the dispersion relation of the free model: $k^2\mapsto k^2+ \epsilon \eta(k)$, with $\epsilon\ll1$ and $\eta(\cdot)$ a suitable function. 
Using this deformed dispersion in \eqref{eq:M6} one can ensure that $\mathcal{M}_6$ coincides with $\mathcal{N}_6$ in any finite volume $L$ and hence Condition~\ref{eq:condition1} is always fulfilled for finite $L$. Both strategies (i) and (ii), however, necessarily produce second-order charges where the coefficient of the six-fermion term ($g_{f;2}^{(6)}({\boldsymbol k})$ in Eq.~\eqref{eq:Qf2nd}) has singularities in momentum space. In real space, these singularities translate into a charge density that does not decay, i.e.\ the second-order charges are non-local.

In summary, restricting to zeroth order charges with a quadratic term, Condition~\ref{eq:condition1} is necessary and sufficient for the system \eqref{eq:FermionicH} to have a complete set of charges with densities that decay in space at second order. We remark that if one changes the dispersion relation, for instance by considering a system on the lattice, the situation becomes richer and will be discussed in a separate work \cite{bertini2022wip}.

Let us now characterise the solutions to Condition~\ref{eq:condition1}. We have the following~\cite{Note2}
\begin{property} 
\label{p:prop1}
The only potentials fulfilling Condition~\ref{eq:condition1} and admitting a power-series expansion around 0 are
\be
V_{a b}(k) = a\left( 1- \sqrt{b} k \coth \sqrt{b} k\right),\quad a,b\in \mathbb R. 
\label{eq:sol}
\ee
\end{property}
Restricting to $b>0$ (we seek potentials that are well defined for all $k\in\mathbb R$) we see that that \fr{eq:sol} correspond to the inverse-sinh-squared Calogero-Sutherland potential up to mass and momentum rescaling~\cite{sutherland2004beautiful}. In particular we find the two following limiting cases 
\be
\lim_{\substack{b\to 0 \\ ab = - 3 \gamma}}V_{ab}(k) = \gamma k^2,\quad \lim_{\substack{b\to\infty \\ a = -\gamma/\sqrt{b}}}V_{ab}(k) = \gamma |k|. 
\label{eq:CSCalogero}
\ee
The first is nothing but the Fourier transform of the integrable Cheon-Shigehara potential \eqref{eq:Etiennespotential} at order $O(\beta)$ when the regulator is removed, while the second is the inverse-squared Calogero-Sutherland potential~\cite{sutherland2004beautiful}. We have conducted a numerical check of Condition~\ref{eq:condition1} for several classes of singular potentials but have failed to find any additional solutions. 

We note that to the best of our knowledge, \eqref{eq:sol} and \eqref{eq:CSCalogero} correspond to the only known integrable potentials for \eqref{eq:FermionicH} in the thermodynamic limit. Importantly both cases give rise to theories with a single species of quasi-particles (smoothly connected to free fermions for vanishing $V$). For the $k^2$ potential one can use a standard result of Fourier analysis
 ~\cite{hormander2015analysis, strichartz2003guide}
 to construct a complete set of ``ultra-local" charges \cite{korepin1993quantum} with density supported on a single point. To this end it is sufficient to take the set of charges constructed choosing $f(k)\in\{k^{2n}\}_{n=1}^{\infty}$.

Given the set of conserved charges \eqref{eq:charges} we can now define the quasi-particle number operator $n(k,x)$. Our starting point are the
operatorial densities $q_{f}(x)$ of the charges \eqref{eq:charges}, which can be chosen with at least a power-law decaying density if Condition~\ref{eq:condition1} is fulfilled. 
A convenient choice is 
\begin{align}
&q_{f}(x)= \frac{1}{L}\! \sum_{k_1,k_2} f(k_1)e^{i x (k_2-k_1)} \psi^\dag_{k_1}\! \psi^{\phantom{\dag}}_{k_2}\label{eq:density}\\
&\quad+\frac{1}{L^2}\! \sum_{{\boldsymbol k},k_4}  (g_{f;1}^{(4)}(\boldsymbol {k})+g_{f;2}^{(4)}(\boldsymbol {k})) e^{i x (k_4+k_3-k_2-k_1)} \psi^\dag_{k_1}\!\!\cdots\psi^{\phantom{\dag}}_{k_4}
\notag\\
&\quad+\frac{1}{L^3}\!\! \sum_{{\boldsymbol k},k_6} g_{f;2}^{(6)}({\boldsymbol k}) e^{i x (k_6+k_5+k_4-k_3-k_2-k_1)}\psi^{\dag}_{k_1}\!\!\cdots \psi^{\phantom{\dag}}_{k_6}\notag
\end{align}
corrected by higher orders in $V$. Note that this choice of density operator is \emph{not} Hermitian but one can straightforwardly recover an Hermitian density via $q_{f}(x)\mapsto (q_{f}(x)+q^\dag_{f}(x))/2$.

To construct $n(k,x)$ we consider linear combinations of $\{q_{f}(x)\}$ with smooth $f$'s and select the contributions of a single non-interacting mode $k$. The linear combination is then chosen such that we obtain the density associated with $f(p)=\delta_{k,p}$. Since \eqref{eq:density} is linear in $f$ this is always possible if $\{f(k)\}$ is a complete set in $L^2(\mathbb R)$. Specifically we propose the following definition  
\be
n(k,x):= \frac{L}{2\pi}\frac{\partial }{\partial f(k)} q_{f}(x)\,,
\ee
where $k$ obeys free quantisation conditions. This operator is characterised by the following properties:

\noindent (i) The moments of $n(k,x)$ are the densities
\be
\frac{2\pi}{L}\sum_k f(k) n(k,x) = q_{f}(x). 
\ee

As shown below this implies that the thermodynamic limit of the expectation value of $n(k,x)$ on a translationally invariant state gives the density of quasi-particles.

\noindent (ii) $n(k,x)$ is conserved, i.e., it fulfils 
\begin{align}
\label{ghdeq}
&\partial_t  n(k,x) + \partial_x  j_n(k,x)=0\ ,\nn
&j_n(k,x):= \frac{L}{2\pi} \frac{\partial}{\partial f(k)} j_{f}(x)\ ,
\end{align}
where
$j_{f}(x)$ is the current associated with $q_{f}(x)$ via the continuity equation~\cite{Note2}
\be
\partial_t   q_{f}(x)+\partial_x  j_{f}(x)=0.
\ee

\noindent (iii) An explicit expression of $n(k,x)$ can be obtained from \eqref{eq:density}. We note that for $V=0$ the resulting expression differs from the free fermion Wigner operator, but the latter can be easily recovered by modifying the definition of the densities of the conserved charges.

If the moments of $n(k,x)$ are sufficiently local in space \footnote{Quasi-local moments definitely have this property, but moments that decay as power-laws in space are likely to be sufficient~\cite{bulchandani2021quasiparticle,arsenio2012boltzmann}.} then (i) and (ii) allow us to interpret Eq.~\eqref{ghdeq} as an operatorial progenitor of the GHD equation~\eqref{ghd}. Property (iii) makes the relation between GHD and the BBGKY hierarchy explicit. The expectation value of $n(k,x)$, which fulfils the GHD equation \eqref{ghd}, is recovered as a specific sum of cumulants fulfilling the BBGKY hierarchy. The expansion of $\braket{n(k,x)}$ up to order ${\cal O}(V^n)$ involves cumulants of order $2n+2$. Therefore, in order to recover the full GHD equation one needs the \emph{entire} BBGKY hierarchy. 

Our previous discussion implies that, at second order, $n(k,x)$ is sufficiently local only if $V({\boldsymbol k})$ is taken to be the Cheon-Shigehara or Calogero-Sutherland potential. Remarkably, however, at first order the moments are quasi-local for \textit{any} smooth potential $V({\boldsymbol k})$, integrable or not~\footnote{{\color{blue}}This observation could be related to Ref.~\cite{brandino2015glimmers}}. This suggests that in the weak-interaction regime GHD physics is robust against non-integrable perturbations. Namely, having a weak non-integrable potential in \eqref{eq:FermionicH}  does not preclude the formulation of the BBGKY hierarchy in terms of a GHD equation at first order in perturbation theory, but only at second order. This means that GHD will be applicable on a longer time scale than naively expected and perhaps partially explains the relevance of GHD in modelling actual experiments~\cite{schemmer2019generalized,malvania2020generalized, bouchoule2022generalized}.

In order to fully make contact with GHD, the thermodynamic limit of the expectation values of $n(k,x)$ and ${j}_n(k,x)$ in an energy eigenstate should match the known formulae for Thermodynamic Bethe ansatz (TBA) integrable models. Namely, at order $m$ in perturbation theory one should have
\begin{align}
\braket{n(k,x)}\bigl |_{m} &=\rho(k)+O(V^{m+1}), \label{eq:root}\\
\braket{j_n(k,x)}\bigl|_m &=v(k)\rho(k)+O(V^{m+1}),\label{eq:currentroot}
\end{align}
where $\rho(k)$  is the quasi-particle density in momentum space and $v(k)$ is the group velocity of stable particle and hole excitations around the energy eigenstate~\footnote{$x|_m$ indicates that $x$ is truncated at order $m$ in $V$}. For a given state both these quantities depend non-trivially on the two body interactions characterising the integrable model~\cite{bertini2016transport, castro2016emergent, borsi2021current, cubero2021form}. More precisely, they are determined by two integral equations involving the two-particle ``scattering phase shifts"~\cite{takahashi1999thermodynamics}. Upon assuming local equilibration~\cite{alba2021generalizedhydrodynamic}, the relations \eqref{eq:root} and \eqref{eq:currentroot} allow one to go from the operatorial continuity equation \eqref{ghdeq} to the GHD equation~\eqref{ghd}. 

One can verify that for ${m=1}$ Eqs.~\eqref{eq:root} and \eqref{eq:currentroot} are fulfilled for any potential. Namely, using $\braket{\psi^\dag_p \psi^{\phantom{\dag}}_q}\equiv\delta_{p,q}\vartheta(q)$ and $\braket{\psi^\dag_p \psi^{{\dag}}_q}=0$, we find at first order~\cite{Note4}
\begin{align}
\!\!\!\!\!\braket{n(k,x)} &\!\!=\! \frac{\vartheta(k)}{2\pi}\!\left[1+\!\!\int\!\! {\rm d}q\,  K(k-q)\vartheta(q)\right]+
O(V^2),\label{eq:pertcharge}\\
\!\!\!\!\!\braket{j_n(k,x)} &\!\!=\! \frac{\vartheta(k)}{\pi}\!\left[k+\!\!\int\!\! {\rm d}q\,  K(k-q)q \vartheta(q)\!\right]+
O(V^2),
\label{eq:pertcurr}
\end{align}
where $K(k)=\partial_k({V(k)}/{k})$. These agree with the first order expansion of $\rho(k)$ and $v(k)$ in an integrable model with a single species of quasi-particles and a scattering phase shift $V(k)/k$ (perturbatively small). The latter is then interpreted as  the scattering phase shift of the effective integrable model describing \eqref{eq:FermionicH} at first order. In particular, using the potentials \eqref{eq:sol} and \eqref{eq:CSCalogero} we recover the first order expansion of the known TBA expressions of $\rho(k)$ and $v(k)$ in the Lieb-Liniger and Calogero-Sutherland models~\cite{Note2}. The extension of these results to higher orders in the integrable case will be reported elsewhere \cite{bertini2022wip}. 

We note that at first order in $V$ the above programme can be generalized to the bosonic case, i.e.\ when the operators $\psi$ entering the Hamiltonian \eqref{eq:FermionicH} satisfy canonical commutation relations. We are again able to construct charges for any potential, but now the expectation value of these charges in an eigenstate is in general divergent at first order in $V$, as reported in Sec.\ref{sec:boson} of the SM.

\textit{Discussion.} In this Letter we showed how to systematically
derive a dissipationless Boltzmann equation for certain ``dressed" quasi-particles from the BBGKY hierarchy for weakly interacting many-particle systems and derived an explicit expression for the density of the corresponding mode occupation operator. In order to enable a GHD description this density must have good locality properties, which we find to be the case precisely for the known integrable potentials. This suggests that our ``integrability condition" provides an exact characterisation of all integrable systems (even those not solvable by Bethe ansatz) with a single species of quasi-particles. Our construction further suggests that in weakly interacting models the GHD description is robust against non-integrable perturbations, which is relevant for applications of GHD to cold-atom experiments. Our work can be extended in a number of directions. Firstly, the case of integrable models with several species of stable quasi-particles can be analysed in a similar way, but a number of interesting complications occur \cite{bertini2022wip}. Secondly, from our ``operatorial" GHD equation it should be possible to derive corrections to GHD~\cite{fagotti2017higher, fagotti2020locally, denardis2022correlation, ruggiero2020quantum, alba2021generalizedhydrodynamic} in a systematic fashion. Finally, our approach can be used to study the effects of general weak integrability breaking interactions~\cite{doyon2017a, friedman2020diffusive, durnin2021nonequilibrium, lopez2021hydrodynamics, bastianello2021hydrodynamics, bertini2016thermalization, bertini2015prethermalization, bastianello2019generalized}. An important goal is to investigate how, and over what time-scales, they render perturbed GHD descriptions invalid.

{\em Acknowledgments:}
This work has been supported by the Royal Society through the University Research Fellowship No.\ 201101 (BB) and the EPSRC under grant EP/S020527/1 (EG and FHLE).

\bibliography{./bibliography}

\onecolumngrid
\newpage 

\newcounter{equationSM}
\newcounter{figureSM}
\newcounter{tableSM}
\stepcounter{equationSM}
\setcounter{equation}{0}
\setcounter{figure}{0}
\setcounter{table}{0}
\makeatletter
\renewcommand{\theequation}{\textsc{sm}-\arabic{equation}}
\renewcommand{\thefigure}{\textsc{sm}-\arabic{figure}}
\renewcommand{\thetable}{\textsc{sm}-\arabic{table}}

\begin{center}
{\large{\bf Supplemental Material for\\
``BBGKY Hierarchy and Generalised Hydrodynamics"}}
\end{center}

Here we report some useful information complementing the main text. In particular
\begin{itemize}
\item[-] In Section~\ref{sec:PerturbativeCharges} we explicitly compute the charges at the first two orders in perturbation theory. 
\item[-] In Section~\ref{sec:proofp1} we prove Property~\ref{p:prop1}. 
\item[-] In Section~\ref{sec:commutation} we show that the charges \eqref{eq:charges} commute up to second order.
\item[-] In Section~\ref{sec:current} we present an explicit expression for the current $j_{f,x}$ associated to the density $q_{f,x}$ in Eq.~\eqref{eq:density}.
\item[-] In Section~\ref{sec:TBA} we present an explicit Thermodynamic Bethe ansatz derivation of \eqref{eq:pertcharge} and \eqref{eq:pertcurr} in the cases of Lieb--Liniger and Calogero--Sutherland.  
\end{itemize}

\section{Perturbative Determination of the Charges}
\label{sec:PerturbativeCharges}

\subsection{First Order}

The first order correction to the charge, $Q_{f;1}$, is obtained by solving the following equation
\be
[H_0, Q_{f;1}]  = - [H_1, Q_{f; 0}] \,,
\label{eq:chargePT1st}
\ee
where $Q_{f; 0}$ is given in Eq.~\eqref{eq:zerothQ}. The commutator on the r.h.s.\ is immediately evaluated as 
\be
[H_1, Q_{f; 0}] = \frac{1}{L}\sum_{k_1, \ldots, k_4} V(k_1,k_2,k_3,k_4)(f(k_1)+f(k_2)-f(k_3)-f(k_4)) \psi_{k_1}^\dag \psi^{\dag}_{k_2}\psi_{k_3}^{\phantom{\dag}} \psi^{\phantom{\dag}}_{k_4}. 
\label{eq:genericcommutator}
\ee
Therefore making the ansatz 
\be
Q_{f;1}= \frac{1}{L}\sum_{k_1, \ldots, k_4} h_{f;1}^{(4)}(k_1,\ldots, k_4) \psi_{k_1}^\dag \psi^{\dag}_{k_2}\psi_{k_3}^{\phantom{\dag}} \psi^{\phantom{\dag}}_{k_4}
\label{eq:Qf1stapp}
\ee
we see that \eqref{eq:chargePT1st} is solved if 
\be
h_{f;1}^{(4)}(k_1,\ldots, k_4) (k_1^2+k_2^2-k_3^2-k_4^2) = V(k_1,k_2,k_3,k_4)(f(k_4)+f(k_3)-f(k_2)-f(k_1))\,.
\ee
The latter equation admits solutions only if the r.h.s\ vanishes when the l.h.s.\ does. This means that we should have 
\be
V(k_1,k_2,k_3,k_4)(f(k_1)+f(k_2)-f(k_3)-f(k_4))=0\qquad (k_1,k_2,k_3,k_4)\in \mathcal K_{4}\,,
\label{eq:constraint}
\ee
where we set 
\be
\mathcal K_{4} = \{ (k_1,k_2,k_3,k_4):\, k_j\in \frac{2\pi}{L}\mathbb Z,\quad k_1^2+k_2^2-k_3^2-k_4^2=0\}.
\ee
First we note that if 
\be
k_1+k_2-k_3-k_4\neq0
\ee
the condition \eqref{eq:constraint} is immediately satisfied. Indeed $V(k_1,k_2,k_3,k_4)$ vanishes because of the momentum-conserving Kronecker Delta in Eq.~\eqref{eq:VW}. Therefore, the non-trivial condition is  
\be
V(k_1,k_2,k_3,k_4)(f(k_1)+f(k_2)-f(k_3)-f(k_4))=0\qquad (k_1,k_2,k_3,k_4)\in \mathcal M_{4}\,,
\label{eq:constraint2}
\ee
where 
\be
\mathcal M_{4} = \{ (k_1,k_2,k_3,k_4):\, k_j\in \frac{2\pi}{L}\mathbb Z,\quad k_1^2+k_2^2-k_3^2-k_4^2=0,\quad  k_1+k_2-k_3-k_4=0\}.
\label{eq:MEP2}
\ee
It is easy to see that \eqref{eq:constraint2} is immediately satisfied. Indeed, by solving explicitly the constraints and writing $k_1^2+k_2^2-k_3^2-(k_1+k_2-k_3)^2=-2(k_3-k_1)(k_3-k_2)$, the manifold $\mathcal M_{4}$ can be rewritten as 
\be
\mathcal M_4 = \{ (k_1,k_2,k_2,k_1):\, k_j\in \frac{2\pi}{L}\mathbb Z\}\cup\{ (k_1,k_2,k_1,k_2):\, k_j\in \frac{2\pi}{L}\mathbb Z\},
\label{eq:MEP2simp}
\ee 
and it is immediate to see that 
\be
(f(k_1)+f(k_2)-f(k_3)-f(k_4))=0,\qquad (k_1,k_2,k_3,k_4)\in \mathcal M_{4}\,.
\ee
Therefore \eqref{eq:chargePT1st} is solved by \eqref{eq:Qf1st} with the choice 
\be
h_{f;1}^{(4)}(k_1,\ldots, k_4) = V(k_1,k_2,k_3,k_4)\frac{f(k_1)+f(k_2)-f(k_3)-f(k_4)}{k_4^2+k_3^2-k_2^2-k_1^2},\qquad (k_1,k_2,k_3,k_4)\in \mathcal M_{4}. 
\ee 
Using the momentum-conserving delta function in $V(k_1,k_2,k_3,k_4)$ we can then write 
\be
Q_{f;1}= \frac{1}{L}\sum_{k_1, \ldots, k_3} g_{f;1}^{(4)}(k_1,\ldots, k_3) \psi_{k_1}^\dag \psi^{\dag}_{k_2}\psi_{k_3}^{\phantom{\dag}} \psi^{\phantom{\dag}}_{k_1+k_2-k_3}
\label{eq:Qf1stappmom}
\ee
where we introduced
\be
g_{f;1}^{(4)}(k_1,\ldots, k_3) =V(k_1,k_2,k_3,k_1+k_2-k_3)\frac{f(k_1)+f(k_2)-f(k_3)-f(k_1+k_2-k_3)}{(k_1+k_2-k_3)^2+k_3^2-k_2^2-k_1^2},
\label{eq:g1}
\ee
for $k_1^2+k_2^2-k_3^2-(k_1+k_2-k_3)^2\neq0$, extended by continuity at $k_3=k_1$ and $k_3=k_2$.

\subsection{Second Order}

Repeating the procedure we find that the second order correction to the charge, $Q_{f;2}$, has to fulfil the following condition 
\be
[H_0, Q_{f;2}]  = - [H_1, Q_{f;1}].
\label{eq:chargePT2nd}
\ee
In particular, using the quartic form of the potential and the expression \eqref{eq:Qf1st} for the first order correction to the charge we find 
\be
 [H_1, Q_{f;1}] = [H_1, Q_{f;1}]_4 + [H_1, Q_{f;1}]_6,
\ee
where $[H_1, Q_{f;1}]_4$ and $[H_1, Q_{f;1}]_6$ denote respectively quartic and sextic terms in the fermionic operators, their explicit expression reads as 
\begin{align}
[H_1, Q_{f;1}]_4 &=  \frac{2}{L}\sum_{k_1,\cdots, k_4} c_{f,2}^{(4)}(k_1, \cdots, k_4) \psi^\dag_{k_1} \psi^{\dag}_{k_2}\psi^{\phantom{\dag}}_{k_3} \psi^{\phantom{\dag}}_{k_4}\notag\\
[H_1, Q_{f;1}]_6 &=  \frac{4}{L^2}\sum_{k_1,\cdots, k_6} c_{f,2}^{(6)}(k_1, \cdots, k_6) \psi^\dag_{k_1} \psi^{\dag}_{k_2}\psi^{\dag}_{k_3}\psi^{\phantom{\dag}}_{k_4}\psi^{\phantom{\dag}}_{k_5} \psi^{\phantom{\dag}}_{k_6}\,.
\end{align}
where we set 
\begin{align}
&c_{f,2}^{(4)}(k_1, \cdots, k_4) := \frac{1}{L}\sum_q \left( V(k_1, k_2, k_1+k_2-q, q) h_{f;1}^{(4)}(q,k_1+k_2-q,k_3, k_4)\right.\notag\\
&\qquad\qquad\qquad\qquad\qquad\qquad\left.- h_{f;1}^{(4)}(k_1, k_2, k_1+k_2-q,q) V(q, k_1+k_2-q,k_3, k_4)\right)\notag\\
&\qquad\qquad\qquad\quad= \frac{\delta_{k_1+k_2,k_3+k_4}}{L}\sum_q  V(k_1, k_2, k_1+k_2-q, q)V(q, k_3+k_4-q,k_3, k_4)\left(\frac{\Delta f(k_{1},k_{2},q)}{\Delta \epsilon(k_{1},k_{2},q)}-\frac{\Delta f(k_{3},k_{4},q)}{\Delta \epsilon(k_{3},k_{4},q)}\right)
\notag\\
\notag\\
&c_{f,2}^{(6)}(k_1, \cdots, k_6) := \delta_{k_1+k_2+k_3,k_4+k_5+k_6} \mathcal A_{k_1 k_2 k_3}\mathcal A_{k_4 k_5 k_6}\left[V(k_4-k_1)V(k_3-k_5)\left(\frac{\Delta f(k_{5},k_6,k_{3})}{\Delta \epsilon(k_{5},k_6,k_{3})}-\frac{\Delta f(k_{1},k_{2},k_4)}{\Delta \epsilon(k_{1},k_{2},k_4)}\right)\right]
\notag\\
\notag\\
&\Delta f(x,y,z) := f(x)+f(y)-f(z)-f(x+y-z), \quad \Delta \epsilon(x,y,z) := x^2+ y^2 - z^2-(x+y-z)^2,
\end{align}
and $A_{a b c}[\cdot]$ is the operator that antisymmetrises with respect to $a$, $b$, and $c$.  This means that $Q_{f;2}$ has to have the same structure, namely 
\begin{align}
Q_{f;2} =&  \frac{1}{L}\sum_{k_1,\cdots, k_4} h_{f;2}^{(4)}(k_1, k_2, k_3,k_4) \psi^\dag_{k_1} \psi^{\dag}_{k_2}\psi^{\phantom{\dag}}_{k_3} \psi^{\phantom{\dag}}_{k_4}\notag\\
&+ \frac{1}{L^2}\sum_{k_1,\cdots, k_6} h_{f;2}^{(6)}(k_1, k_2, k_3 , k_4, k_5,k_6) \psi^\dag_{k_1} \psi^{\dag}_{k_2}\psi^{\dag}_{k_3}\psi^{\phantom{\dag}}_{k_4}\psi^{\phantom{\dag}}_{k_5} \psi^{\phantom{\dag}}_{k_6}\,.
\label{eq:ansatzQ2nd}
\end{align}
In order for \eqref{eq:ansatzQ2nd} to solve \eqref{eq:chargePT2nd} we have to impose  
\begin{align}
  h_{f;2}^{(4)}(k_1, k_2, k_3,k_4) (k_1^2+k_2^2-k_3^2-k_4^2) &= -2c_{f,2}^{(4)}(k_1, \cdots, k_4)\,,\notag\\
  h_{f;2}^{(6)}(k_1, k_2, k_3, k_4,k_5,k_6) (k_1^2+k_2^2+k_3^2-k_4^2-k_5^2-k_6^2) &= -4c_{f,2}^{(6)}(k_1, \cdots, k_6)\,.
\end{align}
As before, it is easy to see that these equations admit solutions only if 
\begin{align}
c_{f,2}^{(4)}(k_1, \cdots, k_4) &= 0 \qquad (k_1, \cdots, k_4)\in \mathcal M_{4},\label{eq:quartic}\\
c_{f,2}^{(6)}(k_1, \cdots, k_6) &= 0 \qquad (k_1, \cdots, k_6)\in \mathcal M_{6},\label{eq:sextic}
\end{align}
where $\mathcal M_{4}$ is defined in \eqref{eq:MEP2} and we introduced 
\be
\mathcal M_{6} = \{ (k_1,k_2,k_3,k_4,k_5,k_6):\, k_j\in \frac{2\pi}{L}\mathbb Z,\quad k_1^2+k_2^2+k_3^2-k_4^2-k_5^2-k_6^2=0,\quad  k_1+k_2+k_3-k_4-k_5-k_6=0\}.
\label{eq:MEP3}
\ee
It is easy to see that \eqref{eq:quartic} is satisfied for any two-body potential. Indeed, first we note that $ \mathcal M_{4}$ can be expressed as in Eq.~\eqref{eq:MEP2simp} and then we observe 
\begin{align}
& c_{f,2}^{(4)}(k_1,k_2,k_{2/1},k_{1/2})  = \sum_q  |V(k_1, k_2, k_1+k_2-q, q)|^2 \left(\frac{\Delta f(k_1,k_2,q)}{\Delta \epsilon(k_1,k_2,q)}-\frac{\Delta f(k_{1/2},k_{2/1},q)}{\Delta \epsilon(k_{1/2},k_{2/1},q)}\right)=0\,
\end{align}
where we used 
\be
V(k_1, k_2, k_3, k_4) = V(k_4, k_3, k_2, k_1)^*\,.
\ee
To treat \eqref{eq:sextic} we note 
\begin{align}
 c_{f,2}^{(6)}(k_1,\cdots,k_6)  = & -\mathcal A_{k_1 k_2 k_3}\mathcal A_{k_4 k_5 k_6}\!\!\left[\frac{V(k_4-k_1)V(k_3-k_5)}{\Delta \epsilon(k_{1},k_{2},k_4)}\right] \Delta_3 f(k_1,\ldots,k_5)\,,\quad k_j\in\mathcal M_{6}
\label{eq:constraintapp}
\end{align}
where we used 
\be
\Delta \epsilon(k_5,k_6,k_3)=\Delta \epsilon(k_{1},k_{2},k_4)\,,\quad k_j\in\mathcal M_{6}\,,
\ee
and defined 
\be
\Delta_3 f(k_1, k_2, k_3, k_4, k_5) : = f(k_1)+f(k_2)+f(k_3)-f(k_4)-f(k_5)-f(k_1+k_2+k_3-k_4-k_5)\,.
\ee
Furthermore, we note that the latter quantity vanishes identically on the sub-manifold 
\be
\mathcal N_{6} = \bigcup_{\substack{i=1,2,3 \\ j=4,5,6}} \{(k_1,k_2,k_3,k_4,k_5,k_6)\in \mathcal M_{6}: \quad k_i=k_j\}\,.
\ee
Indeed, it is easy to see that 
\be
\mathcal N_{6} = \bigcup_{\sigma\in S_3} \{(k_1,k_2,k_3,k_{\sigma(1)},k_{\sigma(2)},k_{\sigma(3)}):\, k_j\in \frac{2\pi}{L}\mathbb Z\}\,,
\ee
where $\sigma$ is a permutation and $S_3$ is the group of permutations of three objects. 

This means that in order for \eqref{eq:sextic} to hold for generic functions $f$ one has to impose
\be
\mathcal A_{k_1 k_2 k_3}\mathcal A_{k_4 k_5 k_6}\!\!\left[\frac{V(k_4-k_1)V(k_3-k_5)}{\Delta \epsilon(k_{1},k_{2},k_4)}\right] = 0, \qquad \forall  k_i\in\mathcal M_{6}\setminus\mathcal N_{6}\,.
\label{eq:keypropertyapp}
\ee
This is nothing but Condition~\ref{eq:condition1} of the main text. If this condition holds we can write 
\begin{align}
Q_{f;2} =&  \frac{1}{L}\sum_{k_1,\cdots, k_3} g_{f;2}^{(4)}(k_1, k_2, k_3) \psi^\dag_{k_1} \psi^{\dag}_{k_2}\psi^{\phantom{\dag}}_{k_3} \psi^{\phantom{\dag}}_{k_1+k_2-k_3}\notag\\
&+ \frac{1}{L^2}\sum_{k_1,\cdots, k_5} g_{f;2}^{(6)}(k_1, k_2, k_3 , k_4, k_5) \psi^\dag_{k_1} \psi^{\dag}_{k_2}\psi^{\dag}_{k_3}\psi^{\phantom{\dag}}_{k_4}\psi^{\phantom{\dag}}_{k_5} \psi^{\phantom{\dag}}_{k_1+k_2+k_3 - k_4 - k_5}\,,
\label{eq:Q2ndmom}
\end{align}
with 
\begin{align}
 g_{f;2}^{(4)}(k_1, k_2, k_3)  &= -\frac{2 c_{f,2}^{(4)}(k_1, \cdots, k_1+k_2-k_3)}{(k_1^2+k_2^2-k_3^2-(k_1+k_2-k_3)^2)}\,,\label{eq:g24}\\ 
 g_{f;2}^{(6)}(k_1, k_2, k_3, k_4,k_5) &= - \frac{4 c_{f,2}^{(6)}(k_1, \cdots, k_1+k_2+k_3-k_4-k_5)}{(k_1^2+k_2^2+k_3^2-k_4^2-k_5^2-(k_1+k_2+k_3-k_4-k_5)^2)}\label{eq:g26}
\end{align}
extended by continuity for $(k_1,k_2,k_3,k_1+k_2-k_3)\in \mathcal M_{4}$ and $(k_1,k_2,k_3,k_4,k_5,k_1+k_2+k_3-k_4-k_5)\in \mathcal M_{6}$ respectively. 

\section{Proof of Property~\ref{p:prop1}}
\label{sec:proofp1}

In this section we prove Property~\ref{p:prop1}. We begin by noting that $\mathcal M_{6}$ can be parametrised as 
\begin{align}
\mathcal M_{6} = &\{ (k_1,k_2,k_3,k_4,s_+(k_1,k_2,k_3,k_4),s_-(k_1,k_2,k_3,k_4)):\, \Delta(k_1,k_2,k_3,k_4)>0\}\notag\\
&\cup\{ (k_1,k_2,k_3,k_4,s_-(k_1,k_2,k_3,k_4),s_+(k_1,k_2,k_3,k_4)):\, \Delta(k_1,k_2,k_3,k_4)>0\},
\label{eq:MEP3simp}
\end{align}
where we introduced 
\begin{align}
s_\pm(x,y,z,t) &:= \frac{1}{2}\left(x+y+z-t\mp\sqrt{\Delta(x,y,x,t)}\right)\notag\\
\Delta(x,y,z,t)  & : ={x^2-2 x y-2x z+2 x t+y^2-2 y z+2 yt+z^2+2 z t-3 t^2}\,.
\end{align}
Next, we define 
\be
f(k_1,k_2,k_3,k_4)= 
{\prod_{i=1}^3\prod_{j=4}^6(k_i-k_j)}
\mathcal A_{k_1k_2k_3}\mathcal A_{k_4k_5k_6}\!\!\left[\frac{V(k_4-k_1)V(k_3-k_5)}{\Delta \epsilon(k_{1},k_{2},k_4)}\right] \bigr|_{k_5=s_+;k_6=s_-}\,.
\ee
In this new notation the task is to find all potentials written as 
\be
V(k)= \sum_{n\geq 1} c_{2n} k^{2n},
\label{eq:series}
\ee
for which 
\be
\label{eq:condition1rewritten}
f(k_1,k_2,k_3,k_4)= 0, \qquad \forall k_1,k_2,k_3,k_4\in \mathbb R\,.
\ee
To this end, let us consider a necessary condition and seek for the potentials for which 
\be
 h(x) \equiv f(x,2x,-3x,-x)  =0, \qquad \forall x. 
\label{eq:condition1p}
\ee
Explicitly we find  
\begin{equation}
    \begin{aligned}
    h(x)=-8x^7(&-V(3x)V(2x)+V(4x)V(6x)-3V(3x)V(6x)+2V(2x)V(6x)-2V(x)V(3x)+4V(x)V(2x)\\
    &-V(x)^2-V(2x)V(4x)-2V(2x)^2+3V(3x)^2)\,.
    \end{aligned}
\end{equation}
Using 
\be
\tanh(2x)=\frac{2\tanh x}{1+\tanh^2 x}, \qquad x\in \mathbb C,
\ee
one can explicitly verify that for 
\begin{equation}\label{eq:potentialtanhx}
    V_{\alpha\beta}(k)=\alpha\left(1-{\beta k}{\coth (\beta k)} \right)\,,\qquad \alpha,\beta \in \mathbb C,
\end{equation}
Condition \eqref{eq:condition1p} is satisfied. Let us now argue that these are the only potentials with regular power series expansion around 0 fulfilling \eqref{eq:condition1p}. 

First we observe that if $c_2=0$ in \eqref{eq:series} the potential cannot fulfil \eqref{eq:condition1p}. This can be seen as follows. Let $\bar n>1$ be the smallest $n>1$ for which $c_{2n}$ in \eqref{eq:series} does not vanish. Then we have
\be
k_{4\bar n} = \frac{1}{(4\bar n)!} \partial_x^{4\bar n} \frac{h(x)}{(-8x^7)}\bigr |_{x=0}= - 36 \left(2^{2\bar n}-1\right) \left(3^{2\bar n}-1\right) \left(-3^{2\bar n+1}+3\ 2^{2\bar n}+4^{2\bar n}-1\right) c_{2 \bar n}^2\,,
\ee
which never vanishes for $\bar n>1$. Therefore we can assume $c_2\neq0$ and write 
\be
V(k) \mapsto c_2 (k^2+\sum_{n\geq 2} \tilde c_{2n} k^{2n}).
\ee
Next we note 
\begin{equation}\label{gamma}
    k_{2 n+2} = \frac{1}{(2 n+2)!} \partial_x^{2 n+2} \frac{h(x)}{(-8x^7)} \bigr |_{x=0} = (-4+35\cdot 2^{2n}-60\cdot 3^{2n}+32\cdot 4^{2n}-3 \cdot 6^{2n}) \tilde c_{2n}+P_n(\tilde c_{4}, \tilde c_6,..., \tilde c_{2n-2}),
\end{equation}
with $P_n$ a known polynomial. The coefficient in front of $\tilde c_{2n}$ vanishes for $n=1,2$ and never vanishes for $n\geq 3$. Imposing $k_{2 n+2}=0$ for all $n$ then uniquely fixes the values of all $\tilde c_{2n}$'s for $n\geq 3$ in terms of $\tilde c_4$. It follows that, up to a multiplicative constant, there is at most a one-parameter family of solutions to \eqref{eq:condition1p} and, therefore, \eqref{eq:potentialtanhx} are the only possibilities.

To conclude, we have to show that \eqref{eq:potentialtanhx} fulfil also the more general condition \eqref{eq:condition1rewritten}. This is indeed the case as can be seen, for instance, using the symbolic simplification of Mathematica. 

Restricting the family \eqref{eq:potentialtanhx} to real potentials we have that the only solutions of \eqref{eq:condition1rewritten} with regular power series expansion around 0, and hence decaying rapidly enough in real space, are 
\be
V_{ab}(k)=a\left(1- \sqrt b k \coth (\sqrt b k) \right), \qquad a,b\in\mathbb R\,,
\ee
as claimed in the main text.

\section{Commutation of the Charges}
\label{sec:commutation}

In this section we prove that the charges~\eqref{eq:charges} commute up to second order in $V$. Writing explicitly the commutator up to second order we have 
\begin{align}
[ Q_{f}, Q_{g}] =&  [Q_{f;0}, Q_{g;0}] + [Q_{f;1}, Q_{g;0}] +[Q_{f;0}, Q_{g;1}]\notag\\
&+[Q_{f;2},Q_{g;0}] + [Q_{f;1}, Q_{g;1}] +[ Q_{f;0}, Q_{g;2}].
\label{eq:commutator}
\end{align}
The first term on the first line of the l.h.s.\ is clearly 0, while the second and the third are opposite. Concerning the second line we have 
\begin{align}
[Q_{f;2},Q_{g;0}] + [Q_{f;1}, Q_{g;1}] +[ Q_{f;0}, Q_{g;2}] = & \frac{1}{L} \sum_{k_1,k_2,k_3} q_4(k_1,k_2,k_3) \psi^\dag_{k_1} \psi^{\dag}_{k_2}\psi^{\phantom{\dag}}_{k_3} \psi^{\phantom{\dag}}_{k_1+k_2-k_3}\\
&+ \frac{1}{L^2} \sum_{k_1,\cdots,k_5} q_6(k_1,k_2,k_3,k_4,k_5) \psi^\dag_{k_1} \psi^{\dag}_{k_2}\psi^{{\dag}}_{k_3}\psi^{\phantom{\dag}}_{k_4}\psi^{\phantom{\dag}}_{k_5} \psi^{\phantom{\dag}}_{k_1+k_2+k_3-k_4-k_5},\notag
\end{align}
where we introduced   
\begin{align}
q_4(k_1,k_2,k_3) = & \frac{1}{L} \sum_q V(k_1, k_2, k_1+k_2-q, q) V(q,k_1+k_2-q,k_3, k_4) \notag\\
&\qquad\qquad\times\left\{q_4^{(20)}(k_1,k_2,k_3,q)+q_4^{(02)}(k_1,k_2,k_3,q)+q_4^{(11)}(k_1,k_2,k_3,q)\right\},\\
q_4^{(20)}(k_1,k_2,k_3,q) = & 2\frac{\Delta f(k_1,k_2,k_3)}{\Delta \epsilon(k_1,k_2,k_3)}\left\{\frac{\Delta g(k_3,k_1+k_2-k_3,q)}{\Delta \epsilon(k_3,k_1+k_2-k_3,q)}-\frac{\Delta g(k_1,k_2,q)}{\Delta \epsilon(k_1,k_2,q)}\right\},\\
q_4^{(02)}(k_1,k_2,k_3,q) = 
&2\frac{\Delta g(k_1,k_2,k_3)}{\Delta \epsilon(k_1,k_2,k_3)}\left\{\frac{\Delta f(k_1,k_2,q)}{\Delta \epsilon(k_1,k_2,q)}-\frac{\Delta f(k_3,k_1+k_2-k_3,q)}{\Delta \epsilon(k_3,k_1+k_2-k_3,q)}\right\},\\
q_4^{(11)}(k_1,k_2,k_3,q) = &2\left(\frac{\Delta f(k_1,k_2,q)}{\Delta \epsilon(k_1,k_2,q)}\frac{\Delta g(k_3,k_1+k_2-k_3,q)}{\Delta \epsilon(k_3,k_1+k_2-k_3,q)}-\frac{\Delta g(k_1,k_2,q)}{\Delta \epsilon(k_1,k_2,q)}\frac{\Delta f(k_3,k_1+k_2-k_3,q)}{\Delta \epsilon(k_3,k_1+k_2-k_3,q)}\right),
\end{align}
and 
\begin{align}
q_6(k_1,k_2,k_3,k_4,k_5) = & V(k_4-k_1)V(k_3-k_5)\left\{q_6^{(20)}(k_1,\ldots,k_5)+q_6^{(02)}(k_1,\ldots,k_5)+q_6^{(11)}(k_1,\ldots,k_5)\right\}\\
q_6^{(20)}(k_1,k_2,k_3,k_4,k_5) = & 4\frac{\Delta_3 g(k_1,k_2,k_3,k_4,k_5)}{\Delta_3 \epsilon (k_1,k_2,k_3,k_4,k_5)} \left(\frac{\Delta f(k_{1},k_{2},k_4)}{\Delta \epsilon(k_{1},k_{2},k_4)}-\frac{\Delta f(k_{5},k_1+k_2+k_3-k_4-k_5,k_{3})}{\Delta \epsilon(k_{5},k_1+k_2+k_3-k_4-k_5,k_{3})}\right),\\
q_6^{(02)}(k_1,k_2,k_3,k_4,k_5) = & 4\frac{\Delta_3 f(k_1,k_2,k_3,k_4,k_5)}{\Delta_3 \epsilon (k_1,k_2,k_3,k_4,k_5)} \left(\frac{\Delta g(k_{5},k_1+k_2+k_3-k_4-k_5,k_{3})}{\Delta \epsilon(k_{5},k_1+k_2+k_3-k_4-k_5,k_{3})} -\frac{\Delta g(k_{1},k_{2},k_4)}{\Delta \epsilon(k_{1},k_{2},k_4)}\right),\\
q_6^{(11)}(k_1,k_2,k_3,k_4,k_5) = & 4\left(\frac{\Delta g(k_{1},k_{2},k_4)}{\Delta \epsilon(k_{1},k_{2},k_4)}\frac{\Delta f(k_{5},k_1+k_2+k_3-k_4-k_5,k_{3})}{\Delta \epsilon(k_{5},k_1+k_2+k_3-k_4-k_5,k_{3})}\right.\notag\\
&\quad-\left.\frac{\Delta f(k_{1},k_2,k_{4})}{\Delta \epsilon(k_{1},k_2,k_{4})}\frac{\Delta g(k_{5},k_1+k_2+k_3-k_4-k_5,k_{3})}{\Delta \epsilon(k_{5},k_1+k_2+k_3-k_4-k_5,k_{3})}\right),\\
\Delta_3 \epsilon(k_1,\ldots,k_5) = & k_1^2+k_2^2+k_3^2-k_4^2-k_5^2-(k_1+k_2+k_3-k_4-k_5)^2\,.
\end{align}
Noting 
\begin{align}
\Delta f(k_1,k_2,k_3) &=\Delta f(k_1,k_2,q)-\Delta f(k_3,k_1+k_2-k_3,q),\\
\Delta_3 f(k_1,k_2,k_3,k_4,k_5) &=\Delta f(k_1,k_2,k_4)-\Delta f(k_5,k_1+k_2+k_3-k_4-k_5,k_3), 
\end{align}
one can easily verify 
\begin{align}
q_4^{(20)}(k_1,k_2,k_3,q)+q_4^{(02)}(k_1,k_2,k_3,q)+q_4^{(11)}(k_1,k_2,k_3,q)&=0,\qquad \forall (k_1,k_2,k_3,q)\in \mathbb R^4,\notag\\
q_6^{(20)}(k_1,k_2,k_3,k_4,k_5)+q_6^{(02)}(k_1,k_2,k_3,k_4,k_5)+q_6^{(11)}(k_1,k_2,k_3,k_4,k_5)&=0,\qquad \forall (k_1,k_2,k_3,k_4,k_5)\in \mathbb R^5,
\end{align}
for any $f$ and $g$ (even when replacing the dispersion $k^2$ with a generic function $\epsilon(k)$). Therefore \eqref{eq:commutator} vanishes up to $O(V^3)$.

\section{Currents at First Order} 
\label{sec:current}

The current is obtained by substituting the explicit form of the density (cf.~\eqref{eq:density}) into the Continuity Equation \eqref{ghdeq}. In particular, evaluating the derivative we find 
\be
\partial_t q_{f}(x) = i [H_0, q_{f;0}(x)]+i [H_0, q_{f;1}(x)]+ i [H_1, q_{f;0}(x)]. 
\ee
An explicit calculation gives 
\begin{align}
[H_0, q_{f;0}(x)] =&  \frac{1}{L} \sum_{p,q} f(p) (p^2-q^2)  e^{-i x (q-p)}\psi^\dag_{p} \psi^{\phantom{\dag}}_{q}\,,\\
[H_0, q_{f;1}(x)] =&  \frac{1}{L^2} \sum_{k_1,\ldots,k_4}  r_f(k_1,\ldots, k_4)  e^{-i x (k_1+k_2-k_3-k_4)}\psi^\dag_{k_1} \psi^{\dag}_{k_2}\psi^{\phantom{\dag}}_{k_3} \psi^{\phantom{\dag}}_{k_4}\,,\\
[H_1, q_{f;0}(x)] =&  \frac{1}{L^2} \sum_{k_1,\ldots,k_4}  s_f(k_1,\ldots, k_4) e^{-i x (k_1+k_2-k_3-k_4)} \psi_{k_1}^\dag \psi^{\dag}_{k_2}\psi_{k_3}^{\phantom{\dag}} \psi^{\phantom{\dag}}_{k_4},
\end{align}
with 
\begin{align}
r_f(k_1,\ldots, k_4)  &=g_{f;1}^{(4)}(k_1,k_2,k_3) (k_1^2+k_2^2-k_3^2-k_4^2)\\
s_f(k_1,\ldots, k_4) &= (f(k_1+k_2-k_3)-f(k_2) )V(k_1-k_3) + (f({k_1+k_2-k_4})-f(k_1))V(k_4-k_2). 
\end{align}
Noting that 
\begin{align}
\lim_{k_4\to k_1+k_2-k_3}\frac{r_f(k_1,\ldots, k_4)+s_f(k_1,\ldots, k_4)}{(k_1+k_2-k_3-k_4)}=&\left\{2(k_3-k_1-k_2)g_{f;1}^{(4)}(k_1,k_2,k_3)-(V(k_1-k_3)-V(k_2-k_3)) f'(k_3)\right.\notag\\
&+\left. \frac{1}{2}(V'(k_1-k_3)-V'(k_2-k_3))\left[2f(k_3)-f(k_1)-f(k_2)\right]\right\},
\end{align}
we can then define the current operator fulfilling \eqref{ghdeq} as follows
\begin{align}
j_{f;1}(x) =& \frac{1}{L}  \sum_{p,q} f(p) (p+q)  e^{-i x (q-p)}\psi^\dag_{p} \psi^{\phantom{\dag}}_{q} \notag\\ 
&+\frac{1}{L^2} \sum_{k_1,\ldots,k_4} \frac{r_f(k_1,\ldots, k_4)+s_f(k_1,\ldots, k_4)}{k_1+k_2-k_3-k_4} e^{-i x (k_1+k_2-k_3-k_4)} \psi_{k_1}^\dag \psi^{\dag}_{k_2}\psi_{k_3}^{\phantom{\dag}} \psi^{\phantom{\dag}}_{k_4}.
\label{eq:current}
\end{align}

\section{TBA derivation of  (\ref{eq:pertcharge}) and (\ref{eq:pertcurr})}
\label{sec:TBA}

Consider an integrable model treatable by Thermodynamic Bethe ansatz and featuring a single species of quasi-particles. In the thermodynamic limit eigenstates are described by a single root density $\rho(k)$, which characterises the distribution of Bethe roots~\cite{takahashi1999thermodynamics}. The latter is related to the so called filling-function, describing the distribution of Bethe integers, as follows~\cite{takahashi1999thermodynamics} 
\be
\rho(k)=\vartheta(k) \rho_t(k), 
\ee
where $\rho_t(k)$ is the solution of 
\be
\rho_t(k) = \frac{1}{2\pi}+\int {\rm d}q\,  \tilde K(k-q) \vartheta(q)\rho_t(q),
\ee
and $K(k)$ is the derivative of the scattering phase shift. On the other hand, the velocity of elementary excitations is determined by the following integral equation 
\be
v(k)\rho_t(k) = \frac{\epsilon'(k)}{2\pi}+\int {\rm d}q\,  \tilde K(k-q) \vartheta(q)v(q)\rho_t(q),
\ee
where $\epsilon(k)$ is the dispersion relation of the free modes. Here we focus on the choices 
\be
\epsilon(k)=k^2,
\ee
and 
\be
\tilde K(k)= \frac{\beta}{1+\beta^2 k^4/4}, -2\pi \beta \delta(k), -\frac{1}{2}\left( \psi_0(\beta+1+i k/2)+\psi_0(\beta+1-i k/2)-\psi_0(1+i k/2)-\psi(1-i k/2)\right),
\ee
where $\psi_0(k)$ is the Digamma function. These three cases describe respectively the Lieb Liniger (for $c=2/\beta$), the inverse-squared, and the inverse-sinh-squared Calogero-Sutherland models (for $\beta=\lambda-1$)~\cite{sutherland2004beautiful}.  

Fixing the Bethe integers and expanding for small $\beta$ we find 
\begin{align}
\rho(k) & = \frac{\vartheta(k) }{2\pi}\left(1+\int {\rm d}q\,   K(k-q) \vartheta(q)\right)+O(\beta^2)\\
v(k) \rho(k) & = \frac{\vartheta(k)}{\pi}\left(k+\int {\rm d}q\,   K(k-q) \vartheta(q)q\right)+O(\beta^2)
\end{align}
with  
\be
K(k)= {\beta}, -2\pi \beta \delta(k), -\frac{\beta}{2} \left(\frac{4 - k^2 \pi^2 {\rm csch}(k \pi/2)^2}{k^2}\right),
\ee
where we used the identity~\footnote{See, e.g., \href{https://functions.wolfram.com/GammaBetaErf/PolyGamma2/17/02/01/}{https://functions.wolfram.com/GammaBetaErf/\\PolyGamma2/17/02/01/}}
\be
\psi_1(1+i k/2)+\psi_1(1-i k/2) = \frac{4 - k^2 \pi^2 {\rm csch}(k \pi/2)^2}{k^2}.
\ee
These expressions agree with \eqref{eq:pertcharge} and \eqref{eq:pertcurr} upon choosing 
\be
V(k) = \beta k^2, - \pi \beta |k|, -2\beta \left(\frac{\pi k}{2}\coth\left(\frac{\pi k}{2}\right)-1\right). 
\ee
Apart from proportionality and scale factors these are indeed the potentials in \eqref{eq:sol} and \eqref{eq:CSCalogero}.

\section{Bosonic case at first order}
\label{sec:boson}
We consider the Hamiltonian $H=H_0+H_1$ with
\be
H_0=\sum_{p} p^2\phi_p^\dag \phi_{p}\ ,\
H_1=\frac{1}{L}\sum_{{\boldsymbol p}} V({\boldsymbol p}) \phi_{p_1}^\dag \phi_{p_2}^\dag\phi^{\phantom{\dag}}_{p_3}\phi^{\phantom{\dag}}_{p_4},
\label{eq:BosonicH}
\ee
where now the $\phi$s satisfy canonical (bosonic) commutation relations. $V(p)$ is supposed in the form
\be
V(\boldsymbol p) =  \frac{1}{4}\delta_{p_1+p_2,p_3+p_4} \mathcal S_{p_1 p_2}\mathcal S_{p_3 p_4}[V(p_1-p_3)],
\label{eq:VWboso}
\ee
where $\mathcal S_{k_1 \cdots k_n}$ is an operator that acts by symmetrising with respect to $\{k_1, \ldots, k_n\}$. Without loss of generality we can again assume $V(k)$ to be even. However, in general one has $V(0)\neq 0$. Similarly to the fermionic case, we consider the zeroth-order charges
\be
Q_{f;0} =  \sum_{p} f(p) \phi_p^\dag \phi_{p},
\label{eq:zeroth}
\ee
where $f(p)$ is any function, which we take to be smooth. The calculation of the conserved charges at first order in $V$ are similar to the fermionic case and one finds

\be
Q_{f;1}= \frac{1}{L}\sum_{k_1, \ldots, k_3} g_{f;1}^{(4)}(k_1,\ldots, k_3) \phi_{k_1}^\dag \phi^{\dag}_{k_2}\phi_{k_3}^{\phantom{\dag}} \phi^{\phantom{\dag}}_{k_1+k_2-k_3}
\label{eq:Qf1stappmomSM}
\ee
where we introduced
\be
g_{f;1}^{(4)}(k_1,\ldots, k_3) =
\frac{1}{L} (V(k_1-k_3)-V(0))\frac{f(k_1)+f(k_2)-f(k_3)-f(k_1+k_2-k_3)}{(k_1+k_2-k_3)^2+k_3^2-k_2^2-k_1^2},
\label{eq:g1SM}
\ee
for $k_1^2+k_2^2-k_3^2-(k_1+k_2-k_3)^2\neq0$, extended by continuity at $k_3=k_1$ and $k_3=k_2$. However, important changes occur when evaluating such charges in an eigenstate. We find 
\begin{equation}\label{expe}
    \langle n(k,x)\rangle= \frac{\vartheta_b(k)}{2\pi}\left[1+\int {\rm d}q\, K_b(k-q)\vartheta_b(q) \right]\,,
\end{equation}
where
\begin{equation}
    K_b(k)=-\partial_k \frac{V(k)+V(0)}{k}\,,
\end{equation}
and where $\vartheta_b(k)$ is the filling fraction of momenta $k$ in the eigenstate. We note that contrary to the fermionic case this function is \textit{not} bounded and can take arbitrarily large values, since bosons can coincide. But an important difference with the fermionic case is that \eqref{expe} is in general divergent at first order in $V$, since $V(0)\neq 0$ in general. This corresponds to the well-known fact that the expansion in $c$ of energy levels of the Lieb-Liniger model is singular \cite{marino2019exact}.

\end{document}